\documentstyle[dina4p,12pt,epsfig]{article}
\newcommand{\PO}{\em I\! \! P }
\newcommand{\pom}{I\!\!P}
\newcommand{\xpom}{x_{\pom}}

\newcommand{\SCI}{Ingelman_DIS95,Ingelman_LEPTO65,SCIa,SCIb,Ingelman_lishep98}

\newcommand{\BUCHMU}{Buchmueller_DIS95a,Buchmueller_DIS95b,Buchmuller_Hebecker_Mcdermott,Buchmueller_97a,Buchmueller_charm}
\newcommand{\pQCD}{Wusthoff,Diehl1,Diehl2,Bartels_dijet_ws,Bartels_jets,Bartels_asym}

\newcommand{\DVCS}{frankfurt_dvcs3,frankfurt_dvcs1,frankfurt_dvcs2}
\newcommand{\diffpdf}{diff_pdf1,diff_pdf2}

\input feynman
\bigphotons
\begin{document}
\thispagestyle{empty}
\noindent
DESY 98-130                           \hfill ISSN 0418--9833 \\
     \hfill LUNFD6/(NFFL-7157) 1998 \\
%     \hfill Draft no.\ 3 \\
\begin{center}
  \begin{Large}
  \begin{bf}
{Future Diffraction at HERA
\footnote{
To appear in {\em Proc. of the LISHEP workshop on diffractive physics},
  edited by A. Santoro (Rio de Janeiro, Brazil, Feb 16 - 20, 1998)
}}
  \end{bf}
  \end{Large}
\end{center}
\begin{center}
  \begin{large}
H.~Jung\\
    \vspace{0.5cm} 
    {\it Physics Department, Lund University, P.O. Box 118, 221 00 Lund,
Sweden}  
\end{large}
\\
\vspace*{1.cm}
  {\bf Abstract}
\end{center}
\begin{quotation}
\noindent
  {\bf Abstract}
\noindent
{\small

Future prospects of hard diffraction at HERA a reviewed. A selection of processes
which can be calculated in pQCD is given, with emphasis on the separation of
soft and hard diffraction. The main focus will be put on the
energy dependence of diffractive processes
 and signatures for the hard pQCD pomeron. 
Problems in the experimental detection of these processes  and the
expected significance of future measurements at HERA  are discussed.
}
\end{quotation}
%%%%%%%%%%%%%%%%%%%%%%%%%%%%%%%%%%%%%%%%%%%%%%%%%%%%%%%%%%%%%%%%%%%%%%%%%%%%%%
\section{Introduction}
%%%%%%%%%%%%%%%%%%%%%%%%%%%%%%%%%%%%%%%%%%%%%%%%%%%%%%%%%%%%%%%%%%%%%%%%%%%%%%
In a deep inelastic diffractive 
process $ e  p \to e'  p'  X$, where $p'$ represents the  
scattered proton or a low mass final state, and $X$ stands for the 
diffractive hadronic state,  
the cross section can be written as~\cite{Ingelman_Prytz}:    
\begin{equation}
\frac{d^4 \sigma (e p \to e' X p')}{dy\,dQ^2\,dx_{\PO}\,dt}
= \frac{4 \pi \alpha ^2}{y Q^4} 
   \left(   \left( 1 -  y + \frac{y^2}{2} \right) 
                F_2^{D(4)}(x,Q^2;x_{\PO},t)
                   -   \frac{y^2}{2} F_L^{D(4)}(x,Q^2;x_{\PO},t) \right)
\end{equation}
with $y=(q.p)/(e.p)$, $Q^2=-q^2=(e-e')^2$,
$x_{\PO}= (q.\PO)/(q.p) = 1 - (q.p')/(q.p)$ 
and $t=(p-p')^2$,  
where $e$ ($e'$) are the four vectors of the incoming (scattered) electron,
the Bjorken $x$ variable $x= Q^2/(y\cdot s)$ with the total 
center of mass energy $s=(e+p)^2$,
$p$ ($p'$) are the four vectors of the incoming (scattered) proton,  
$q=e-e'$ is
the four vector of the exchanged photon and  
$\PO = p - p'$ corresponds to 
 the four vector of the pomeron, which here only serves as a generic name.
  These variables are 
defined independently of the underlying picture of diffraction.
In analogy to Bjorken-$x$, one can define $\beta = x/x_{\PO}$.
In terms of experimental accessible quantities, these variables can be expressed
as:
\begin{eqnarray}
x_{\PO} & = &\frac{Q^2 + M_X^2}{Q^2 + W^2} \\
\beta & = & \frac{Q^2}{Q^2 + M_X^2} 
\end{eqnarray}
with $M_X$ being the invariant mass of the diffractive ($\gamma^* \PO$)
system and $W$ the mass of the $\gamma^* p$ system.
\par
In the following I shall concentrate on the question of the mechanism
responsible for diffraction. In section 2, I shall briefly describe different
approaches to deep inelastic diffraction and their expected significance. The
question of the energy dependence of diffraction will be addressed. 
In section 3, 
I shall discuss a personal selection of main open questions in diffraction and
in section 4, I will concentrate on the most promising processes suited for a
separation of soft and hard pQCD contributions to diffractive scattering at
HERA. Main emphasis will be put on processes that can be
calculated in pQCD and on the pQCD description of the pomeron.

%%%%%%%%%%%%%%%%%%%%%%%%%%%%%%%%%%%%%%%%%%%%%%%%%%%%%%%%%%%%%%%%%%%%%%%%%%%%%%
\section{Present Knowledge of Deep Inelastic Diffraction at HERA}
%%%%%%%%%%%%%%%%%%%%%%%%%%%%%%%%%%%%%%%%%%%%%%%%%%%%%%%%%%%%%%%%%%%%%%%%%%%%%%
Three different approaches to 
deep inelastic diffractive scattering are mainly discussed in the
literature:
\begin{enumerate}
\item {\bf Resolved pomeron a la Ingelman and Schlein and diffractive parton
densities} 

 In the model of Ingelman-Schlein~\cite{IS}
 $F_2^{D(4)}$ can be written 
 as the product
of the probability, $f_{p\;\PO}$, to find a pomeron  in the proton,
and the
structure function $F_2^{\PO}$ of the pomeron:
\begin{equation}
F_2^{D(4)}(\beta,Q^2;\xpom,t) = f_{p\;\PO}(\xpom,t)  F_2^{\PO}(\beta,Q^2)
\label{F2Ddef}
\end{equation}
In analogy to the quark - parton - model of the proton, 
$\beta$ can be interpreted as the fraction of the 
pomeron momentum carried by the struck quark and
$ F_2^{\PO}(\beta,Q^2)$  can be described in terms of momentum weighted
quark density functions  in the pomeron. 
\par 
Eq.(\ref{F2Ddef}) is a special case
of the more general definition of 
diffractive parton densities \cite{\diffpdf}:
\begin{equation} 
F_2^{D(4)}(\beta,Q^2;\xpom,t)= \sum_i e_i^2 \cdot f^D(\beta,Q^2;\xpom,t)
\end{equation}
where the sum runs over all partons with charge $e_i$.
Here no 
Regge type
factorization of $F_2^{D(4)}$  into 
 a flux, $ f_{p\;\PO}(\xpom,t)$, and $F_2^{\PO}$ is assumed. 
The diffractive parton densities can be subjected to the same DGLAP evolution
equations as used in non - diffractive deep inelastic scattering
\cite{Collins_pom}.

\item {\bf pQCD calculation of diffraction via two gluon exchange} 

The pQCD calculation of $e p \to e' q \bar{q} p'$
 was mainly intended to describe exclusive (or lossless) high $p_T$
di-jet production, and in the model of \cite{Wusthoff} estimates on the
total inclusive diffractive cross section are given.
The calculation of diffractive di-jet production can be
performed in pQCD for large photon virtualities $Q^2$ and 
high $p_T$ of the 
$q (\bar{q})$ 
jets~\cite{\pQCD} or for heavy quarks \cite{Lotter_charm,Diehl_charm}.
\par
Since the processes discussed here are mediated by two gluon exchange, 
different assumptions on the nature of the exchanged gluons 
can be
made: in \cite{Diehl1,Diehl2} 
the gluons are non perturbative, in \cite{Wusthoff}  
they are a hybrid of
non perturbative and perturbative ones and in 
\cite{Bartels_dijet_ws,Bartels_jets} 
they are
taken from a NLO parameterization of the proton structure 
function \cite{GRVa,GRVb}.
The cross section is essentially proportional to the  gluon density 
squared of the proton: 
$\sigma \sim \left[\xpom G_p\left(\xpom,\mu^2 \right)\right]^2 $
at the scale $\mu^2 = p_T^2/(1-\beta)$.
In the case of heavy quarks the cross section is finite for all $p_T$, and
the scale is taken to be
$\mu^2 = (p_T^2 + m_f^2)/(1-\beta)$~\cite{Lotter_charm,Diehl_charm},
 where $m_f$ is the mass of the heavy quark.
 Since the gluon density depends on the scale $\mu^2 $ which is set
 by the details of the interaction, 
 this type of processes violates Regge type factorization.
\par
Due to the different gluon densities, different $\xpom$ dependencies of
the cross sections are expected and are further discussed in
\cite{Bartels_dijet_ws,Bartels_jets}, 
where also numerical estimates are presented.

\item {\bf Semi-classical approach to diffraction and Soft Color Interactions} 

Buchm\"uller et al.\ \cite{\BUCHMU}
attempt to describe $\gamma^* + p \to q + \bar{q} + p'$
and $\gamma^* + p \to q + \bar{q} + g + p'$ in a semi - classical approach where
the partons interact with the color field of the proton.
 The cross section of the first process turns out
to be of similar structure as in the pQCD calculation of~\cite{Bartels_jets}
 and is proportional to a constant, which can be interpreted 
 in the semi-classical approach as the
  gluon density squared of the proton.
 The $q \bar{q} g$ process is
described with a usual boson gluon fusion subprocess
involving 
 an effective diffractive gluon density~\cite{Buchmueller_97a}.
In \cite{hebecker} it is shown that this semi-classical approach is exactly
equivalent to the approach using diffractive parton density functions of quarks
and gluons convoluted with the proper partonic scattering amplitude. Therefore
the semi-classical approach will not be discussed separately in the following.
\par
In the Soft Color Interaction (SCI) approach
events with large rapidity gaps are produced by color
reorientation of the colored partons
originating from the hard interaction process~\cite{\SCI},
 before fragmentation. Also here no pomeron is
explicitly introduced. All parameters in this model are determined by  
non-diffractive deep inelastic scattering, except the probability for soft color
reorientation $R_{SCI}$.
\end{enumerate}

From a phenomenological point of view the 
first approach (resolved pomeron and diffractive parton densities) is the
most advanced. A QCD analysis of
the inclusive structure function $F_2^{D(3)}$ has been
performed
by the H1 collaboration ~\cite{H1_F2D3_97} to describe the $Q^2$ evolution 
including an estimate of the intercept  
of the pomeron trajectory  $\alpha_{\PO}(0)$
 and the parton distribution functions   of the pomeron. 
Here $F_2^{D(3)}$ was described including
 a contribution from sub-leading trajectories, giving a good description of
the data. At the present level of accuracy,  $\alpha_{\PO}(0)$
 is consistent with having no dependence on $Q^2$ \cite{H1_F2D3_97,ZEUS_F2D3}.
for $Q^2 > 1$ GeV$^2$.
\par
It has been shown by J. Collins \cite{Collins_pom} that the QCD factorization
theorem also holds for hard diffraction, saying that
\begin{equation}
F_2^D=\sum_i C_{2i} \otimes f_i^D + \mbox{non-leading power of Q}
\end{equation}
where $\otimes$ indicates the convolution of the 
diffractive parton density, $f_i^D$, 
with
the hard scattering coefficient $C_{2i}$ and the sum is running
 over all partons $i$.
 Therefore standard DGLAP evolution equations
are also applicable to $F_2^D$. 
 These diffractive parton density functions can then
be used to model the hadronic final state, in close analogy to standard
non-diffractive deep inelastic scattering. 
Therefore the full machinery of Monte Carlo
techniques used in deep inelastic scattering can be also applied to 
here~\cite{Jung_lishep98_mc}. 
This approach is very successful in the description of
hadronic final state properties (see for example
\cite{Jung_lishep98_mc,Valkarova_lishep98_had,West_lishep98_charm}).
However
one has to note that if the diffractive cross section can be factorized into
a  diffractive parton density and a hard scattering process. This
implies that  a softer part is left,
which can be identified with a pomeron remnant.
 This remnant does not
participate in the hard interaction and has smaller $p_T$ than the partons of
the hard interaction.
\par 
The proof of factorization given in \cite{Collins_pom} does not imply Regge
factorization, meaning that $\alpha_{\PO}(0)$ found in deep inelastic
diffractive scattering 
needs not to be the same as the one observed in hadron-hadron collisions 
and indeed
measurements of $\alpha_{\PO}(0)$ show that in
deep inelastic diffraction $\alpha_{\PO}(0) \sim 1.2$
\cite{H1_F2D3_97,ZEUS_F2D3}, which is  larger than 
$\alpha_{\PO}(0) = 1.08$ obtained from the total cross section in 
hadron-hadron and photon-hadron
collisions.  
\par
From a theoretical point of view the 
second approach (pQCD calculation of diffraction via two gluon exchange) is
more attractive, since here
diffraction is related to the gluon density squared of the proton,
 and no free parameters are left, except the gluon density.
 Such calculations based on the gluon density squared
  predict a larger
intercept of the pomeron trajectory $\alpha_{\PO}(0)$ than expected from the
soft pomeron. The intercept is essentially
given by the rise of the gluon density at small
values of $x$. Moreover since the gluon density depends also on the scale
$\mu^2$ of the hard subprocess, this calculation predicts a violation of the
Regge type 
factorization  of the cross section into a part which only depends on $\xpom$
and $t$ and another part depending only on $\beta$ and $Q^2$. 
Diffractive processes described by two gluon exchange 
are
not covered by the factorization proof of \cite{Collins_pom}, since all the
partons in the system $M_X$ participate in the hard interaction
and the cross section is found to be of higher twist~\cite{Bartels_jets}.
\par
The pQCD calculations are quite complicated, and presently only the most simple
diagrams have been fully
calculated: $\gamma^* p \to q \bar{q} p$ both for light and
heavy quarks and also the production of vector meson bound states.
Even with only these processes included, an impressively good description of
the hadronic energy flow and vector meson production could be achieved, as is
shown in \cite{Jung_lishep98_mc}.
The contribution for $\gamma^* p \to q g \bar{q} p$ has been estimated
 in a specific
region of the 
phase space, where the gluon has much smaller transverse momentum than
the quarks \cite{Wusthoff,Levin_qqg,Levin_charm}. A more complete
calculation of this process is just being performed \cite{Bartels_qqg}.
\par
Within the present accuracy of diffractive measurements in deep inelastic
scattering at HERA, all three very different approaches to hard diffraction
(resolved pomeron model, pQCD calculation and 
soft color interactions) are able
to describe the experimental data reasonably well.
\begin{figure}[htb]
\begin{center}
\epsfig{figure=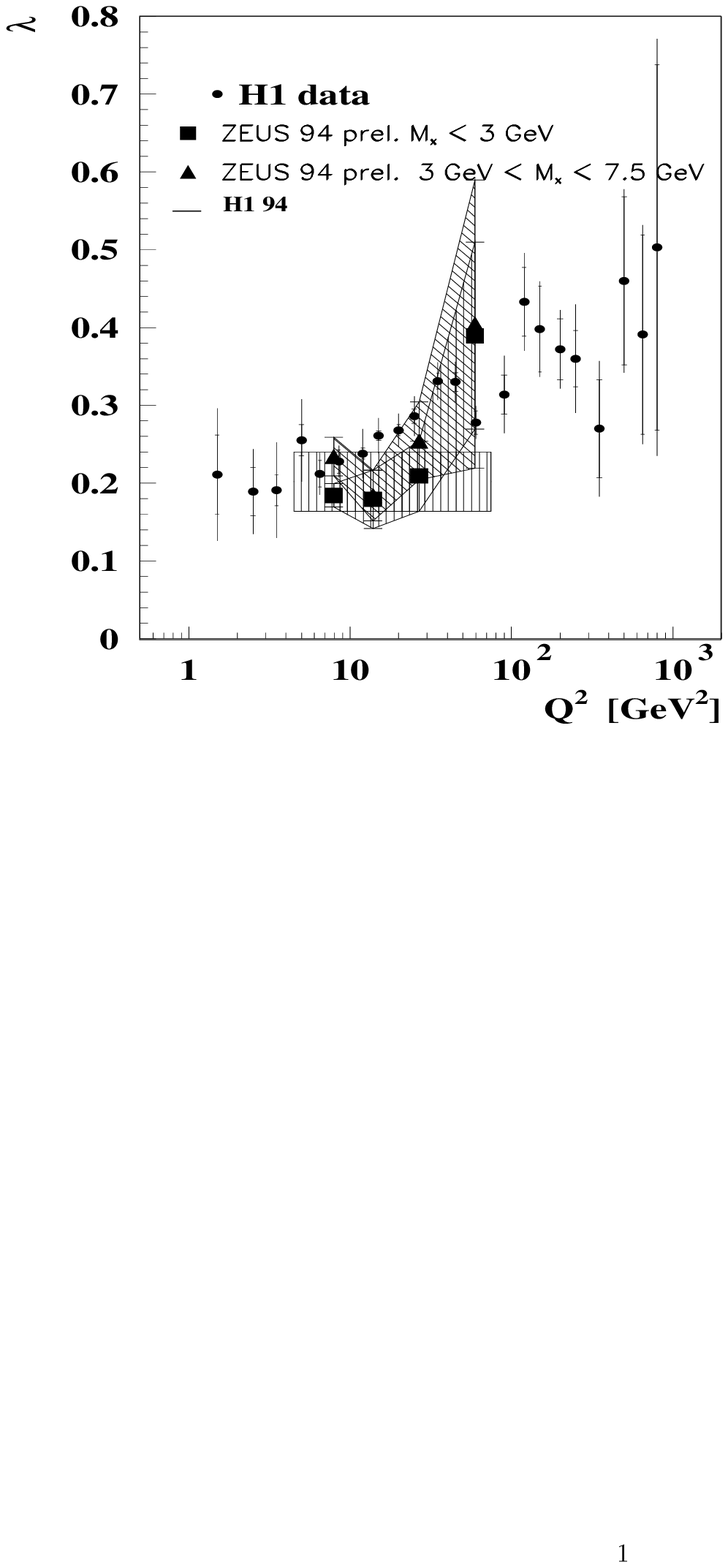,width=12cm,height=10cm,clip=}
\end{center}
\caption{The slope of $F2 \sim  x^{-\lambda}$~\protect\cite{H1_F2} (full dots) 
as a function of $Q^2$
compared to $(\alpha_{\PO}(0) - 1)$ (shaded areas) 
obtained from $F_2^D$~\protect\cite{H1_F2D3_97,ZEUS_F2D3}. The ZEUS data
on diffraction have
been corrected for finite $t$, as given in \protect\cite{ZEUS_F2D3}:
$\alpha_{\PO}(0) = \bar{\alpha_{\PO}} + 0.03$.
\label{F2_slope}}
\end{figure}
\par
It is, however, 
important to understand  the relation of the 
energy dependence of the
inclusive structure function $F_2$ to that of 
the diffractive structure function $F_2^D$. 
The optical theorem relates the total cross section of $\gamma^* p$ to the
forward scattering amplitude of elastic scattering and 
to the diffractive cross section. Writing
the total cross section of   $\gamma^* p$ as
 $\sigma_{tot}(\gamma^* p) \sim x^{-\lambda} $
  and the cross section for diffractive
dissociation  of $\gamma^* p$ as  
$\sigma_{diff}(\gamma^* p) \sim x_{\PO}^{1- 2 \alpha_{\PO}(t)}$
 one can ask whether the
relation $\lambda = (\alpha_{\PO}(0) - 1)$ holds, which would indicate that the
same mechanism (or the same pomeron) is responsible for the rise of the total
inclusive $\gamma p$ cross section at small $x$ and for hard diffractive
scattering. In Fig.~\ref{F2_slope} the exponent $\lambda$ as a function of $Q^2$
as obtained from $F_2$~\cite{H1_F2} is plotted (full dots), 
together with $(\alpha_{\PO}(0) - 1)$
obtained from the measurements of 
$F_2^D$~\cite{H1_F2D3_97,ZEUS_F2D3} (shaded areas). There is
remarkable agreement between the measurement from the total cross section and
from diffraction, although the errors are still large.
The important message is, that $\lambda$ and $\alpha_{\PO}(0) - 1$ is of the 
same magnitude and both are larger than the value obtained from hadron hadron
collisions ($(\alpha_{\PO}(0) - 1)=0.08$). 
This suggests that  the same mechanism  responsible
for the rise of $F_2$ at small $x$ 
is also relevant in deep inelastic diffraction.
%%%%%%%%%%%%%%%%%%%%%%%%%%%%%%%%%%%%%%%%%%%%%%%%%%%%%%%%%%%%%%%%%%%%%%%%%%%%%%
\section{Main open questions}
%%%%%%%%%%%%%%%%%%%%%%%%%%%%%%%%%%%%%%%%%%%%%%%%%%%%%%%%%%%%%%%%%%%%%%%%%%%%%%
One of the main issues to be understood is still the question of the mechanism
responsible for deep inelastic diffraction. If diffraction can be mainly
described by the approach using diffractive parton densities, then a sort of
soft pomeron remnant must be observable. This just follows directly from the
factorization theorem in \cite{Collins_pom}, because not all partons in the
system $M_X$ participate in the hard interaction. On the other hand, in the pQCD
calculation of diffraction via two gluon exchange, all partons participate in
the hard interaction resulting in the absence of a soft pomeron remnant.
 Are there
ways to unambiguously identify a $\PO$ remnant, or on the
contrary is there a significant set 
of events which definitely has no $\PO$ remnant and which can be described by the
pQCD mechanisms outlined above? 
This point is directly related to the question of a separation of hard pQCD
processes in diffraction (where $hard$ means that all partons in the diffractive
system $M_X$ are perturbative) from the part where soft processes are also
involved (identified by the presence of a soft $\PO$ remnant).
\par
Another issue is the energy dependence of
diffraction: from the $F_2^D$ measurements the pomeron intercept is found to be
$\alpha_{\PO} \simeq 1.2$, which is larger than the value found for the soft
pomeron ($\alpha^{soft}_{\PO} \simeq 1.08$). 
Thus the question arises, 
whether the pomeron in deep inelastic scattering is different 
from the one seen in hadron hadron collisions, and  
whether a superposition of a
soft pomeron with a hard QCD pomeron (perturbative two gluon exchange)
is already observed. 
But
the energy dependence of the cross section alone is not sufficient to establish
the existence of a hard pQCD pomeron (two gluon state).
 As in the case of the inclusive
structure function $F_2$ there could be a mixture of soft and hard processes,
resulting in an effective slope $\lambda$ as measured. It is however suggestive,
that the slope of the energy dependence as measured in the inclusive diffractive
structure function $F_2^{D}$ is similar to the one obtained from
 the inclusive structure function $F_2$, from
vector-meson production at large $Q^2$ and from $J/\psi$ production. 
Vector-meson production at large $Q^2$ can be consistently calculated in pQCD, 
at least for longitudinal polarized photons. 
Whereas in the soft pomeron regime shrinkage
of the diffractive peak must be observed (given by $\alpha' = 0.25$), 
A. Levy~\cite{Levy_shrinkage} found evidence for no shrinkage in $J/\psi$
 production
($\alpha' \simeq 0$). If confirmed this would be 
  one of the most important ingredients for a hard
diffractive pQCD process. 
\par
Understanding of diffraction in terms of pQCD requires the separation of 
the soft
from the hard diffractive regime. Several processes have been proposed as
 signatures for hard  pQCD diffraction.
  The most promising processes for hard diffraction accessible by pQCD
 I shall address in the following sections.
%%%%%%%%%%%%%%%%%%%%%%%%%%%%%%%%%%%%%%%%%%%%%%%%%%%%%%%%%%%%%%%%%%%%%%%%%%%%%%
\section{Most promising processes }
%%%%%%%%%%%%%%%%%%%%%%%%%%%%%%%%%%%%%%%%%%%%%%%%%%%%%%%%%%%%%%%%%%%%%%%%%%%%%%
In this section I shall discuss signatures for hard diffraction  
calculable in pQCD:
\begin{itemize}
\item {exclusive di-jets at $Q^2 > 0$ }
\item {charm production }
\item {light vector-mesons at small $t$ and large $Q^2$ or heavy vector-mesons}
\item {vector-mesons at large $t$ }
\item {rapidity gaps between jets}
\item {deep virtual Compton scattering at small and large $t$}
\end{itemize}
All these processes can be calculated completely in pQCD and they all have in
common a specific energy dependence, which is different from the one expected
from soft pomeron processes. Thus the observation and the measurement of these
processes is a crucial test of  pQCD calculations of diffraction, and it will
help solving the question of the relative size and the interplay 
of soft and hard  processes in diffraction. 
\subsection{Di-jets} 
The calculation of exclusive diffractive di-jet production
 $e p \to e' q\bar{q} p$ can be
performed using pQCD for large photon virtualities $Q^2$ and 
high $p_T$ of the 
$q (\bar{q})$ 
jets~\cite{\pQCD}.
Since both quark and anti-quark participate in the hard interaction, they both
receive the same transverse momentum in the $\gamma^* \PO$ system
leaving no remnant behind.
 This has to 
be contrasted to the
approach using diffractive parton densities,
 where also a $q\bar{q}$ state can be produced in a QPM 
process, but there the  quarks have vanishing transverse momentum 
(except from a small intrinsic $p_T$) in the $\gamma^* \PO$ center of mass
system and therefore one of the quarks serves as
a pomeron remnant.
\par
The experimental 
observation of exclusive (or lossless) 
 diffractive di-jet production would give new and
important information, since this process 
is of higher twist nature and it cannot be factorized into a diffractive parton
density convoluted with the hard scattering matrix element. 
Thus it is not covered by the factorization proof of \cite{Collins_pom}. 
\par
\begin{figure}[htb]
\begin{center}
\epsfig{figure=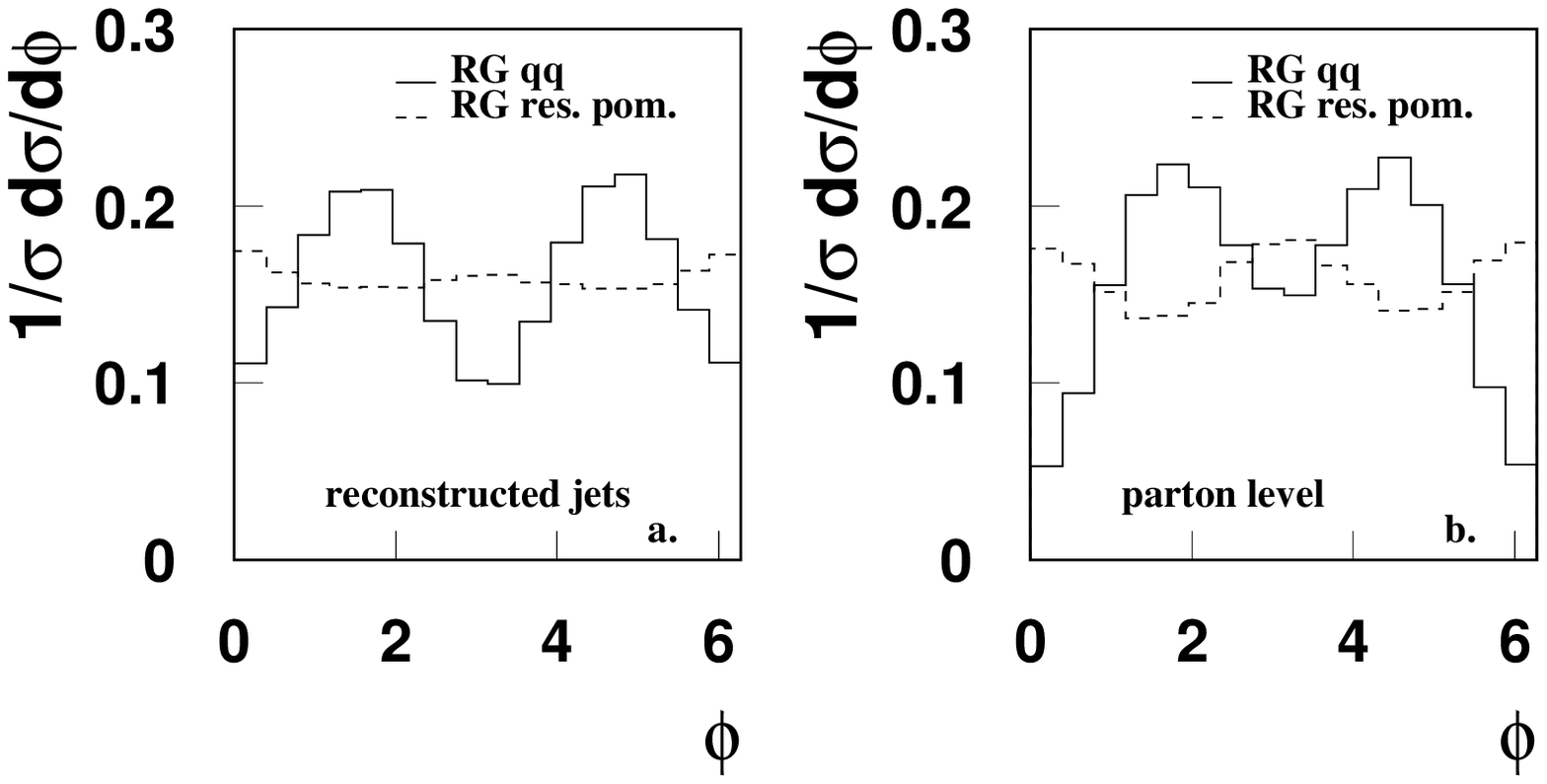,width=18cm,height=11cm}
\end{center}
\caption{$a.$ The $\phi$ dependence of one jet with
respect to the electron plane for 
high $p_T$ di jet events
in the region $0.1 < y < 0.7$, $5 < Q^2< 80$ GeV$^2$
$\xpom < 0.05$ and $p_T^{jet} > 2$ GeV. The solid line shows the 
prediction from the two gluon exchange mechanism
 after jet reconstruction at the hadron level.
The dashed line shows the $\phi$ dependence from a BGF type process in
diffraction (one gluon exchange). In $b.$ the $\phi$ dependence of the 
quark  with the electron plane is shown for comparison.
 The predictions are obtained with the 
 RAPGAP Monte Carlo~\protect\cite{RAPGAP206}.
\label{dijet_phi}}
\end{figure}

The most striking feature of the perturbative QCD calculation of diffractive
$q\bar{q}$ final states is the $\phi$ dependence of
jet production. Here $\phi$ is the angle between the lepton and the quark
plane in the $\gamma^* p$ center of mass system. Since it is
difficult to identify the quark jet at hadron level, 
the jet with the largest $p_T$ can be used
(the partons have the same $p_T$, but  the reconstructed jets not necessarily
because of the jet reconstruction).
The azimuthal asymmetry obtained after jet reconstruction
 is  shown in Fig.~\ref{dijet_phi}, 
where also a comparison with the azimuthal asymmetry expected from a 
diffractive BGF process with one gluon exchange (from a resolved pomeron)
 is given. Also at the hadron level the 
difference between the two approaches is clearly visible.
\par
In the kinematic region defined by $0.1 < y < 0.7$, $5 < Q^2< 80$ GeV$^2$
$\xpom < 0.05$ and $p_T^{jet} > 2$, the cross section for 
$e p \to e' q\bar{q} p$ is 
(calculated with the RAPGAP Monte Carlo using the GRV 
parameterization for the gluon density)
$\sigma^{q\bar{q}} = 46 $ pb, calculated with the RAPGAP Monte Carlo
 using the GRV parameterization for the gluon density.
This should be
compared to the cross section of $\sigma^{res. \PO} = 1138 $ pb
also obtained from RAPGAP, but within the resolved pomeron  model using the 
parameterization of the diffractive parton densities  
by~\cite{H1_F2D3_97}.
However the pQCD calculation of $e p \to e' q\bar{q} p$ is only valid in a
region of relatively small values of $M_X$ or equivalently medium values of 
$\beta > 0.1$. The difficulty of identifying exclusive di-jets is further
discussed in \cite{Jung_lishep98_mc}.
\par
In the region of large $M_X$ the contribution from $q \bar{q} g$ states becomes
important. Estimates of this cross section have been given in 
\cite{Wusthoff,Levin_qqg} in a region where the transverse momentum
 of
the gluon is much 
smaller than that of the quarks. The final state configuration 
 is then similar to the one obtained 
 using boson gluon fusion convoluted with a diffractive gluon density.
  A full
 calculation where no ordering in transverse momentum of the final state partons
 is supposed, is just being performed~\cite{Bartels_qqg}. With this 
 calculation a detailed test of the pQCD prediction for hard diffraction can be
 made: the energy dependence of the cross section is still proportional to the
 gluon density squared, at a scale depending on the transverse momentum of the
 outgoing partons. Thus an energy dependence stronger than expected from soft
 pomeron exchange and a 
  violation of Regge type factorization should be observed.
  Additional information of the underlying mechanism can be obtained from  an
  analysis of the 3 jet final state configuration.
\subsection{Charm - production} 
The calculation of a diffractive $q\bar{q}$ state can also be extended to
heavy quark production \cite{Lotter_charm,Diehl_charm},
where the difficulty of identifying
high $p_T$ di-jets may be avoided by the observation of $D^*$ mesons.  
Because of the heavy quark mass, no $p_T^{cut}$ is 
necessary. 
\begin{figure}[htb]
\begin{center}
\epsfig{figure=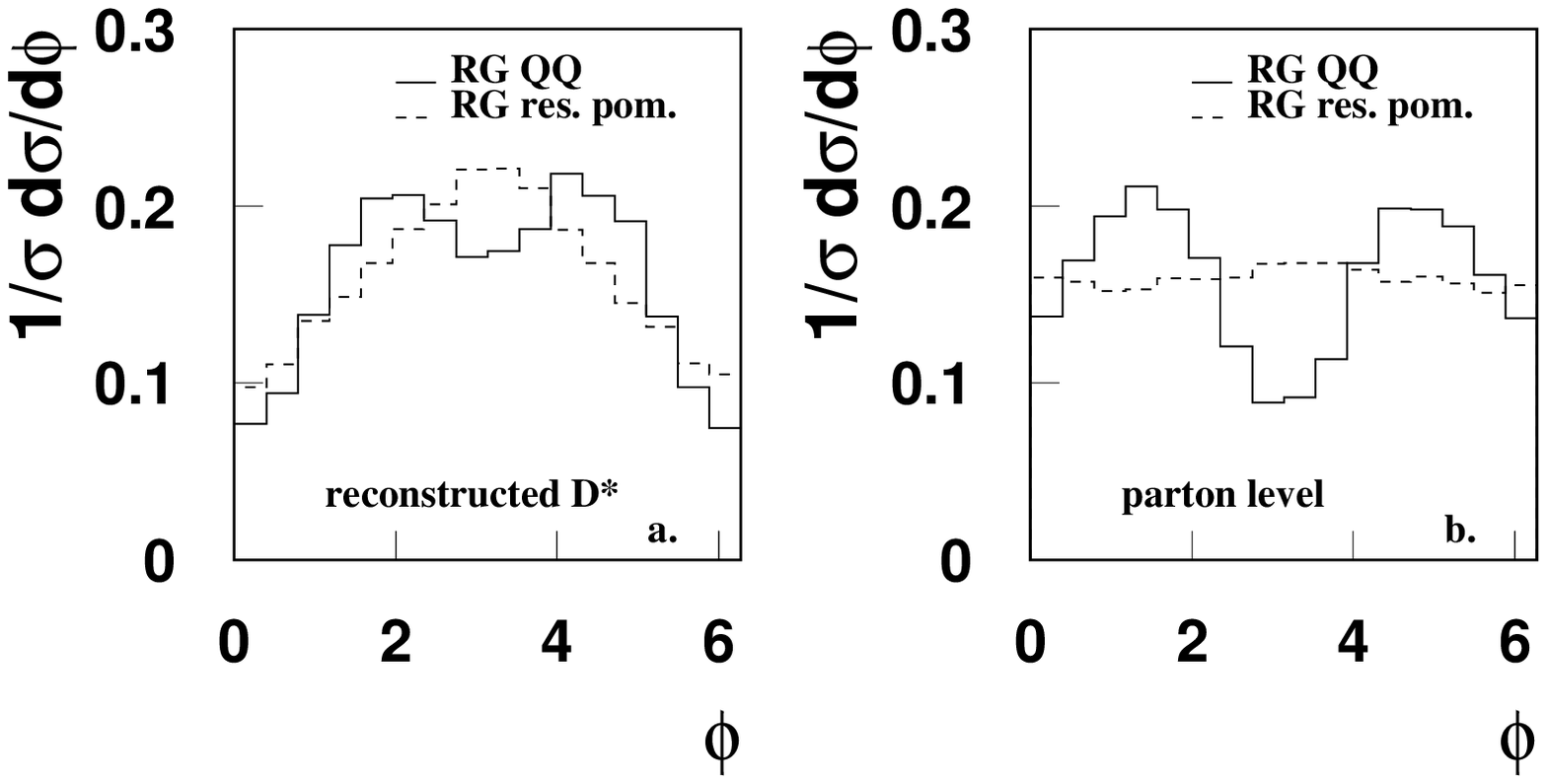,width=18cm,height=11cm}
\end{center}
\caption{$a$. The $\phi$ dependence of the $D^*$  with 
respect to the electron plane 
in the 
kinematic region $0.06 < y < 0.6$, $2 < Q^2< 100$ GeV$^2$
$\xpom < 0.05$, $p_{T\;lab}^{D^*} > 1$ GeV and $|\eta_{lab}^{D^*}| < 1.25$.
The solid line shows the 
prediction from the two gluon exchange mechanism after hadronization.
The dashed line shows the $\phi$ dependence from a BGF type process in
diffraction (one gluon exchange). In $b$ the $\phi$ dependence of the quark
  with respect the electron plane is shown, 
  without the $D^*$ acceptance cuts of $a$.
 The predictions arc obtained with  the 
 RAPGAP Monte Carlo~\protect\cite{RAPGAP206}.
\label{charm_phi}}
\end{figure}

In Fig.~\ref{charm_phi}$a$ the $\phi$ dependence is shown for $D^*$ mesons
 produced
by the two gluon exchange mechanism and compared to the prediction from a
boson gluon fusion process using a diffractive gluon density 
 in a kinematical region typical for
the analyzes of the HERA experiments ($0.06 < y < 0.6$, $2 < Q^2< 100$ GeV$^2$
$\xpom < 0.05$, $p_{T\;lab}^{D^*} > 1$ GeV and $|\eta_{lab}^{D^*}| < 1.25$) 
This process may thus also
be used to differentiate between the two approaches. One should note that the
different $\phi$ distribution observed here, as compared to 
the ones from the jets,
is due to 
the cuts in the laboratory frame used by the experiments to identify the $D^*$
meson. Without the $p_{T\;lab} ^{D^*}$ cut, 
the $\phi$ distribution looks the same as
for the jets. In Fig.~\ref{charm_phi}$b$ the $\phi$ dependence at parton level
 is shown without the $D^*$ acceptance cuts.
\par
Within the search regions of the H1 experiment 
($0.06 < y < 0.6$, $10 < Q^2 < 100$ GeV$^2$, 
$\xpom < 0.05$, $p_T^{D^*} > 1$ GeV and $|\eta^{D^*}| < 1.25$) and 
the ZEUS experiment 
($0.04 < y < 0.7$, $10 < Q^2< 80$ GeV$^2$
$\eta_{max} < 2$,
 $p_T^{D^*} > 1$ GeV and $|\eta^{D^*}| < 1.5$), the 
 cross sections calculated with RAPGAP for the
two gluon exchange process $e p \to e' c\bar{c} p$ including charm
fragmentation into $D^*$ are: $\sigma = 68 $ pb (for H1) and 
$\sigma = 75 $ pb (for ZEUS), compared to the measurement of 
H1 and ZEUS~\cite{West_lishep98_charm}:
$\sigma = 380 \pm ^{150}_{120} \pm  ^{140}_{110}$~pb and 
$\sigma = 875 \pm 248 \pm  ^{395}_{199}$~pb 
obtained from a luminosity of $L = 2.5$~pb$^{-1}$ and  $L = 6.4$~pb$^{-1}$, 
respectively.
The $D^*$ cross section predicted from boson gluon fusion convoluted with the
diffractive gluon density as obtained from a fit to $F_2^{D}$ by H1
\cite{H1_F2D3_97} is: $\sigma = 283.6 $ pb for the H1 measurement and 
$\sigma = 509.1 $ pb for ZEUS measurement.
 Given the large errors on the measurement, no firm conclusion
on the underlying production mechanism can be drawn.
\footnote{In the meantime new results on diffractive charm production have been
presented at ICHEP 98 by the H1 experiment~\cite{H1_diffcharm_ICHEP98},
 with much smaller statistical
error. The cross section for $D^*$ production in the region
$2<Q^2 < 100$ GeV$^2$, $0.05<y<0.7$, $\xpom < 0.04$, 
$p_T^{D^*} > 2$ GeV and $|\eta^{D^*}| < 1.5$
is 
$\sigma= 154 \pm 40 \pm  35$ pb compared to a
prediction from the two gluon exchange mechanism of $\sigma= 112 $ pb 
obtained from
the RAPGAP Monte Carlo program.}
\par
Charm production in deep inelastic diffraction is one of the key processes
for understanding diffraction in terms of pQCD. 
Besides the measurements of the total cross section for diffractive charm
production, the energy (or $\xpom$) dependence will help to differentiate
between different mechanisms. Moreover as shown in Fig.~\ref{charm_phi}, the 
measurement of the $\phi$ dependence is one of the most interesting  ones, since
it allows to unambiguously distinguish the hard pQCD 
process involving 2 gluon exchange from standard boson gluon fusion processes.
However, the cross section is rather small and a large increase in
luminosity is needed for a precise measurement of diffractive charm production
as a function of $\xpom$. It has been argued in~\cite{charm_future},
that a 
luminosity of $L \sim 750 $pb$^{-1}$ is needed for a 
reasonable measurement of the
differential cross section $d \sigma^{D*}/d\xpom$.

\subsection{Vector-meson production} 

The cross section for exclusive vector meson production (light vector-meson
 production at large $Q^2$ and heavy vector meson production even in the
 photo-production region) can be calculated in pQCD, via two gluon exchange,
 similar to the one discussed in the previous sections on high $p_T$ jet and
 open charm production. Measurements 
 (for an overview see \cite{West_lishep98_charm}) 
 of the energy dependence of the 
 vector-meson production cross section are consistent with the pQCD calculations
 and show a much stronger rise with $W$ than expected from soft pomeron
 processes. However only in  photo-production of $J/\psi$ mesons 
 the energy dependence could
 be determined from HERA measurements alone with a reasonable precision.
  It has been shown in
 \cite{vm_future_96} that 10000 events are necessary for a determination of the
 energy slope with an error of $\Delta \alpha_{\PO}(0) \sim 0.01$. 
 In the 1995 data H1 has $\sim$ 100 events for $2 < Q^2 < 8$, thus a factor of 
 100 in luminosity is needed to meet $\Delta \alpha_{\PO}(0) \sim 0.01$. 
 \par
 Even more important for the proof of a pQCD process responsible for 
 vector-meson
 production is the absence of shrinkage of the diffractive peak. In Regge theory
 the $\PO$ trajectory is given by: $\alpha_{\PO} = \alpha_{\PO}(0) +
 \alpha_{\PO}' \cdot t$ with $t$ being the momentum transfer from the proton
 and $\alpha_{\PO}'=0.25$ as determined from soft hadronic collisions.
 A confirmation of the reported evidence for no shrinkage in 
 $J/\psi$ production~\cite{Levy_shrinkage} would be 
 a clear indication of hard pQCD
 processes in heavy vector-meson production. However,  in
 the analysis~\cite{Levy_shrinkage} 
 low energy experiments had to be included.
 To reduce the uncertainty in normalization and background subtraction
 a measurement of $\alpha_{\PO}'$ needs to be done within a single experiment.
 This would require a luminosity of $\sim 250$ pb$^{-1}$
 for a determination of 
 $\Delta \alpha_{\PO}' \sim 0.12$ for $\rho$ production in the range $20 < Q^2 <
 25 $ GeV$^2$, as shown in \cite{vm_future_96}. 
 \par
$J/\psi$ production at large $t$ can be calculated in pQCD, because two large
scales are involved, the $J/\psi$ mass at the photon vertex and the large $t>1$
GeV$^2$ at the proton vertex. For $t \sim m_{J/\psi}^2$ this process can be
calculated using  the BFKL evolution equation.
 Measurements have been performed 
and are in agreement with the pQCD calculations. The
$t$ distribution becomes flatter at large $t$ than expected from an exponential
$t$ distribution. The $W$ dependence of the cross section would yield directly a
measurement of the BFKL pomeron intercept. At present the statistics is
too low for a precise measurement. Besides the energy dependence, 
a measurement of the $t$ slope as a function of $W$ should show again the
characteristic pQCD feature of no shrinkage of the diffractive peak. 
 
\subsection{Rapidity gaps between jets} 

Similarly to $J/\psi$ production at large $t$,
 the cross section for rapidity gaps
between jets can be calculated in pQCD. Instead of the $J/\psi$ mass and the
large $t$ value, here the perturbative scale is set by the transverse momenta of
the jets. This process is  mediated also by two gluon exchange. 
If the rapidity
gap between the jets is large enough, a  rise of the cross section,
typical for BFKL,
should be obtained. Measurements~\cite{Cox_lishep98_rapgap} show that a 
 faction of $\sim 10 \%$ of the jet events have a rapidity gap of
$\Delta \eta > 3.5$. This fraction is a factor of 10 larger than 
in a similar search at $ p\bar{p}$ collisions.
\par
 However one has to worry
about the gap survival probability because of soft interactions between the
remnants of the photon and the proton.
The difference in the fraction of events with rapidity gaps between the jets
as  measured in $p \bar{p}$ collisions and in $e p$ scattering might be
understood in terms of color transparency as argued in \cite{diff_ws_sum}.
Partons in a spatially small configuration can screen each
other's color leading to color transparency and a small interaction cross
section with no final state interaction. On the other hand large size
configurations will have large cross sections and final state interactions.
Hadron hadron interactions are of the latter type, having a large cross section.
In resolved photon processes in photo-production
the final state interactions will fill the gap between jets like in hadron
hadron collisions, resulting in  a smaller cross section for events with
rapidity gaps between the jets. In contrary direct photon processes  
 should yield a larger cross
section, because no final state interaction will spoil the gap. 
It would be important to measure the fraction of events with large rapidity gap
between the jets as a function of $x_{\gamma}$, the fraction of the photon
momentum carried in the hard scattering process.
Even more
 if the same
measurement is performed in deep inelastic scattering ($Q^2 > 0$), where
 resolved
photon processes are less important, a even  higher fraction of events with
rapidity gaps between the jets could be expected than observed in
photo-production and $p \bar{p}$ scattering.
\par
The limited detector acceptance also limits the
size of the rapidity gap that can be observed in the experiments. Especially
BFKL effects could contribute to 
large rapidity gap values. 
This very interesting process has experimental limitations coming from the jet
requirement. A similar measurement, but not relying on jets has been proposed in
\cite{Cox_lishep98_rapgap,Cox_Forshaw}.

\subsection{Deep - Virtual Compton Scattering} 
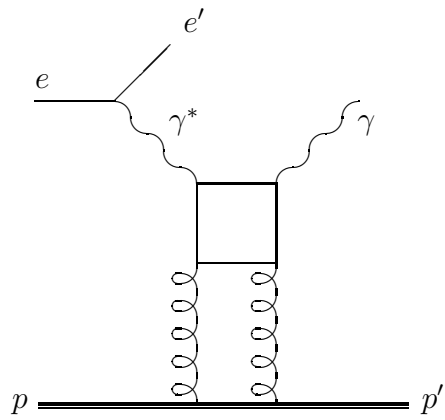
\begin{figure} [htb]
%%%%%%%%%%%%%%%%%%%%%%%%%%%%%%%%%%%%%%%%%%%%%%%%%%%% 
% begin picture basic DIF eq --> e' q' 
%%%%%%%%%%%%%%%%%%%%%%%%%%%%%%%%%%%%%%%%%%%%%%%%%%%% 
\begin{center}
\begin{picture}(15000,20000) 
% draw electron 
\drawline\fermion[\W\REG](3000,15000)[3000] 
\global\advance\pmidy by 500 
\put(\pbackx,\pmidy){$e$} 
% draw photon line 
\drawline\photon[\SE\REG](\fermionfrontx,\fermionfronty)[5] 
\global\advance\pmidx by 500 
\global\advance\pmidy by 500 
\put(\pmidx,\pmidy){$\gamma^*$} 
% draw scattered e 
\drawline\fermion[\NE\REG](\fermionfrontx,\fermionfronty)[3000] 
\global\advance\pbackx by 500 
\global\advance\pbacky by 500 
\put(\pbackx,\pbacky){$e'$} 
%draw scattered q 
\drawline\fermion[\E\REG](\photonbackx,\photonbacky)[3000] 
\global\Xthree = \photonbackx
\global\Ythree = \photonbacky
\global\Xtwo = \fermionbackx
\global\Ytwo = \fermionbacky
\drawline\photon[\NE\REG](\Xtwo,\Ytwo)[5]
\global\advance\pmidx by 1500 
\global\advance\pmidy by 500 
\put(\pmidx,\pmidy){$\gamma $} 
\drawline\fermion[\S\REG](\fermionbackx,\fermionbacky)[3000]
\global\advance\pmidy by 500 
%\put(\pmidx,\pmidy){$q'$} 
%draw initial q 
\drawline\fermion[\S\REG](\Xthree,\Ythree)[3000] 
\global\advance\pmidx by -700 
\global\advance\pmidy by 300 
%\put(\pmidx,\pmidy){$q$} 
\global\Yone = \pbacky 
\global\Xone = \pbackx 
% draw pomeron remnant 
\drawline\fermion[\E\REG](\Xone,\Yone)[3000] 
\global\advance\pmidy by 300 
%\put(\pmidx,\pmidy) {$\bar{q}$} 

% exchange Pomeron 
\drawline\gluon[\S\REG](\Xone,\Yone)[5] 
\global\advance\Xone by 3000 
\drawline\gluon[\S\REG](\Xone,\Yone)[5] 
\global\advance\Xone by -3000 
\global\advance\pbackx by -3000 
% 
% proton vertex 
% 
%\put(\pbackx,\pbacky){\circle*{600}} 
\global\advance\pbacky by - 2500 
%\put(\pbackx,\pbacky){$ (a) $} 
\global\advance\pbacky by + 2500 
\global\advance\pbacky by -100 
\multiput(\pbackx,\pbacky)(0,100){3}{\line(-1,0){6000}}   
\global\Xthree = \pbackx 
\global\Ythree = \pbacky 
\global\advance\Xthree by - 7000 
\put(\Xthree,\Ythree){$p$} 
\multiput(\pbackx,\pbacky)(0,100){3}{\line(1,0){8000}} 
\global\advance\pbackx by + 8500 
\put(\pbackx,\pbacky){$p'$} 
\end{picture}
\end{center}
\caption{Basic diagram for deep virtual compton scattering.
\label{dvcs}}
\end{figure}
Deep virtual compton scattering ($\gamma^* p \to \gamma p'$) is another
example of a diffractive process, which can be calculated in pQCD
\cite{\DVCS}.
 The virtual
photon splits into a $q \bar{q}$ pair which then interacts via two gluon
exchange with the proton, similarly to elastic vector-meson production, but
instead of a vector-meson, a real photon appears in the final state 
(Fig~\ref{dvcs}). 
This process
is again proportional to the gluon density squared of the proton. The main
advantage of this process over vector-meson production is that it can be fully
calculated in pQCD, whereas in the case of vector-meson production a main
uncertainty comes from the poorly known
wave-function of the vector-meson and possible 
relativistic corrections \cite{Jung}.
A more detailed calculation showed that diffractive virtual compton scattering
is sensitive to the off diagonal gluon density, because a finite momentum
transfer is needed to put the incoming virtual photon on mass shell
\cite{\DVCS}.
At large $Q^2$ the energy dependence of this process is expected to be similar
to the ones of vector-meson production, and much stronger than expected from 
soft
pomeron exchange. 
\par
This process would then complete the measurement of vector-meson production at
large energies: production of real photons, $\rho$, $\omega$, $\phi$ and
$J/\psi$. Very interesting would be the measurement of the $t$ slope and 
whether
and how it changes with the final state vector-meson. Again, if this process 
can be
described in terms of pQCD, no shrinkage of the diffractive peak should be
observed ($\alpha_{\PO}' \sim 0$). 
\par
Going one step further, similar to heavy vector-meson production at large $t$,
deep virtual compton scattering 
at large $t$ can be studied. This would then also be
similar to processes with rapidity gaps between jets where one jet is 
replaced by
the real photon in the final state. Because of the appearance of the photon in
the final state, no final state QCD interactions are present, which could spoil
the gap, compared to the measurement with rapidity gaps between jets.
The main
advantage here is, that no jets, nor vector-meson reconstruction
 are required, and that the
 detection of
a high energetic photon  is much simpler, and it can even be detected at much
smaller angles, leading to a larger rapidity region between the proton
dissociative system and the photon. As argued in the previous sections, having
the largest possible rapidity range would be promising for the search of new
small $x$ dynamics like BFKL.
%%%%%%%%%%%%%%%%%%%%%%%%%%%%%%%%%%%%%%%%%%%%%%%%%%%%%%%%%%%%%%%%%%%%%%%%%%%%%%
\section{Summary and Outlook}
%%%%%%%%%%%%%%%%%%%%%%%%%%%%%%%%%%%%%%%%%%%%%%%%%%%%%%%%%%%%%%%%%%%%%%%%%%%%%%
The main problems to be understood in deep inelastic diffraction are the
relatively large diffractive cross section and its energy dependence, which is
stronger than expected from soft processes. The energy dependence might be
understood in terms of pQCD calculations involving 2 gluon exchange processes.
 Such calculations are
consistent with present HERA data, but a firm conclusion on the mechanism
responsible for deep inelastic diffraction cannot be drawn yet.
\par
A significant increase in luminosity is needed for  precise measurements of the
energy dependence and the $t$ slope in various processes in order to study the
contribution from two gluon exchange mechanisms. If these are established
experimentally,
it would be  a major step forward in understanding diffraction in terms of
fully calculable pQCD processes. Even more, this will improve our
understanding of the structure of the proton significantly. 
Such a major increase in luminosity can be expected after the luminosity upgrade
at HERA, which is planned for the year 2000.
\par
Given the importance of understanding diffraction in 
terms of pQCD, one should not forget
the attractive and unique possibility for future experiments 
measuring collisions between electrons from a possible linear collider with
protons from HERA
(500 GeV $e$ $\times$ 820 GeV $p$). 
In such a scenario diffraction and the
structure of the proton could be studied at values of 
$\xpom$ or $x$ a order of
magnitude smaller than presently accessible at HERA. This could open a completely
new area in diffraction. 
%%%%%%%%%%%%%%%%%%%%%%%%%%%%%%%%%%%%%%%%%%%%%%%%%%%%%%%%%%%%%%%%%%%%%%%%%%%%%%
\section*{Acknowledgements}
%%%%%%%%%%%%%%%%%%%%%%%%%%%%%%%%%%%%%%%%%%%%%%%%%%%%%%%%%%%%%%%%%%%%%%%%%%%%%%
It was a great pleasure to participate at this interesting workshop. I am
grateful to the organizers A. Santoro and A. Brandt for this lively workshop and
the stimulating atmosphere. I am grateful to 
J. Bartels, J. Dainton,M. Erdmann, J. Gayler, G. Ingelman and L. J\"onsson for 
careful reading of the manuscript.
%%%%%%%%%%%%%%%%%%%%%%%%%%%%%%%%%%%%%%%%%%%%%%%%%%%%%%%%%%%%%%%%%%%%%%%%%%%%%

% now the references. delete or change fake bibitem. delete next three
%   lines and directly read in your .bbl file if you use bibtex.
\bibliographystyle{prsty} 
\bibliography{habilit}

\begin{thebibliography}{10}

\bibitem{Ingelman_Prytz}
G. Ingelman and K. Prytz, Z. Phys. {\bf C58},  285  (1993).

\bibitem{IS}
G. Ingelman and P. Schlein, Phys. Lett. {\bf B} {\bf 152},  256  (1985).

\bibitem{diff_pdf1}
A. Berera and D. Soper, Phys. Rev. {\bf D} {\bf 50},  4328  (1994).

\bibitem{diff_pdf2}
A. Berera and D. Soper, Phys. Rev. {\bf D} {\bf 53},  6162  (1996),
  hep-ph/9509239.

\bibitem{Collins_pom}
J. Collins, Phys. Rev. {\bf D} {\bf 57},  3051  (1998).

\bibitem{Wusthoff}
M. Wusthoff, Photon diffractive dissociation in deep inelastic scattering,
  1995, \mbox{PhD thesis}, \mbox{DESY-95-166}.

\bibitem{Diehl1}
M. Diehl, Diffraction in electron - proton collisions, 1996, \mbox{PhD thesis}.

\bibitem{Diehl2}
M. Diehl, Z. Phys. {\bf C} {\bf 76},  499  (1997), hep-ph/9610430.

\bibitem{Bartels_dijet_ws}
J. Bartels {\it et~al.},  in {\em Proc. of the Workshop on Future Physics at
  HERA}, edited by A. \mbox{De Roeck}, G. Ingelman, and R. Klanner (DESY,
  Hamburg, 1996), hep-ph/9609239.

\bibitem{Bartels_jets}
J. Bartels, H. Lotter, and M. W\"usthoff, Phys. Lett. {\bf B} {\bf 379},  239
  (1996), hep-ph/9602363.

\bibitem{Bartels_asym}
J. Bartels, C. Ewerz, H. Lotter, and M. Wusthoff, Phys. Lett. {\bf B} {\bf
  386},  389  (1996), hep-ph/9605356.

\bibitem{Lotter_charm}
H. Lotter, Phys. Lett. {\bf B} {\bf 406},  171  (1997), hep-ph/9612415.

\bibitem{Diehl_charm}
M. Diehl, Eur. Phys. J. {\bf C} {\bf 1},  293  (1998), hep-ph/9701252.

\bibitem{GRVa}
M. Gl\"uck, E. Reya, and A. Vogt, Z. Phys. {\bf C} {\bf 53},  127  (1992).

\bibitem{GRVb}
M. Gl\"uck, E. Reya, and A. Vogt, Phys. Lett. {\bf B} {\bf 306},  391  (1993).

\bibitem{Buchmueller_DIS95a}
W. Buchm\"uller and A. Hebecker,  in {\em Proc. of the Workshop on Deep
  Inelastic Scattering and QCD}, edited by J. Laporte and Y. Sirois (Paris,
  April 24 - 28, 1995).

\bibitem{Buchmueller_DIS95b}
W. Buchm\"uller and A. Hebecker, Phys. Lett. {\bf B} {\bf 355},  573  (1995),
  \mbox{DESY 95-077}.

\bibitem{Buchmuller_Hebecker_Mcdermott}
W. Buchmuller, M. McDermott, and A. Hebecker, Nucl. Phys. {\bf B} {\bf 487},
  283  (1997), hep-ph/9607290.

\bibitem{Buchmueller_97a}
W. Buchmuller, M. McDermott, and A. Hebecker, Phys. Lett. {\bf B} {\bf 410},
  304  (1997), hep-ph/9706354.

\bibitem{Buchmueller_charm}
W. Buchmuller, M. McDermott, and A. Hebecker, Phys. Lett. {\bf B} {\bf 404},
  353  (1997), hep-ph/9703314.

\bibitem{hebecker}
A. Hebecker, Diffractive parton distributions in the semiclassical approach,
  1997, hep-ph/9702373.

\bibitem{Ingelman_DIS95}
A. Edin, G. Ingelman, and J. Rathsman,  in {\em Proc. of the Workshop on Deep
  Inelastic Scattering and QCD}, edited by J. Laporte and Y. Sirois (Paris,
  April 24 - 28, 1995).

\bibitem{Ingelman_LEPTO65}
G. Ingelman, A. Edin, and J. Rathsman, Comp. Phys. Comm. {\bf 101},  108
  (1997).

\bibitem{SCIa}
A. Edin, G. Ingelman, and J. Rathsman, Phys. Lett. {\bf B} {\bf 366},  371
  (1996).

\bibitem{SCIb}
A. Edin, G. Ingelman, and J. Rathsman, Z. Phys. {\bf C} {\bf 75},  57  (1997),
  hep-ph/9605281.

\bibitem{Ingelman_lishep98}
G. Ingelman,  in {\em Proc. of the LISHEP workshop on diffractive physics},
  edited by A. Santoro (Rio de Janeiro, Brazil, Feb 16 - 20, 1998).

\bibitem{H1_F2D3_97}
\mbox{H1} Collaboration; C. Adloff~et al., Z. Phys. {\bf C} {\bf 76},  613
  (1997).

\bibitem{ZEUS_F2D3}
\mbox{ZEUS} Collaboration; M. Derrick~et al.,   (1998), \mbox{DESY}-98-084.

\bibitem{Jung_lishep98_mc}
H. Jung,  in {\em Proc. of the LISHEP workshop on diffractive physics}, edited
  by A. Santoro (Rio de Janeiro, Brazil, Feb 16 - 20, 1998).

\bibitem{Valkarova_lishep98_had}
A. Valkarova,  in {\em Proc. of the LISHEP workshop on diffractive physics},
  edited by A. Santoro (Rio de Janeiro, Brazil, Feb 16 - 20, 1998).

\bibitem{West_lishep98_charm}
L. West,  in {\em Proc. of the LISHEP workshop on diffractive physics}, edited
  by A. Santoro (Rio de Janeiro, Brazil, Feb 16 - 20, 1998).

\bibitem{Levin_qqg}
E. Levin and M. Wusthoff, Phys. Rev. {\bf D} {\bf 50},  4306  (1994).

\bibitem{Levin_charm}
E. Levin, A. Martin, M. Ryskin, and T. Teubner, Z. Phys. {\bf C} {\bf 74},  671
   (1997).

\bibitem{Bartels_qqg}
J. Bartels,  in {\em Proc. of the LISHEP workshop on diffractive physics},
  edited by A. Santoro (Rio de Janeiro, Brazil, Feb 16 - 20, 1998).

\bibitem{H1_F2}
\mbox{H1} Collaboration; C.~Adloff {\it et~al.}, Nucl. Phys. {\bf B} {\bf 497},
   3  (1997), hep-ex/9703012.

\bibitem{Levy_shrinkage}
A. Levy, Phys. Lett. {\bf B} {\bf 424},  191  (1998).

\bibitem{RAPGAP206}
H. Jung, {\em \mbox{T}he RAPGAP Monte Carlo for Deep Inelastic Scattering,
  version 2.06}, Lund University, 1998,
  \verb+http://www-h1.desy.de/~jung/rapgap.html+.

\bibitem{H1_diffcharm_ICHEP98}
\mbox{H1} Collaboration; T. Ahmed~et al.,  in {\em Proc. of the 29th
  International Conference on High Energy Physics, ICHEP 98} (Vancover, Canada,
  July, 1998), contributed paper 566.

\bibitem{charm_future}
A. Mehta, J. Phillips, and B. Waugh,  in {\em Proc. of the Workshop on Future
  Physics at HERA}, edited by A. \mbox{De Roeck}, G. Ingelman, and R. Klanner
  (DESY, Hamburg, 1996), p.\ 704.

\bibitem{vm_future_96}
H. Abramowicz {\it et~al.},  in {\em Proc. of the Workshop on Future Physics at
  HERA}, edited by A. \mbox{De Roeck}, G. Ingelman, and R. Klanner (DESY,
  Hamburg, 1996), p.\ 635.

\bibitem{Cox_lishep98_rapgap}
B. Cox,  in {\em Proc. of the LISHEP workshop on diffractive physics}, edited
  by A. Santoro (Rio de Janeiro, Brazil, Feb 16 - 20, 1998).

\bibitem{diff_ws_sum}
H. Abramowicz, J. Bartels, L. Frankfurt, and H. Jung,  in {\em Proc. of the
  Workshop on Future Physics at HERA}, edited by A. \mbox{De Roeck}, G.
  Ingelman, and R. Klanner (DESY, Hamburg, 1996), p.\ 635.

\bibitem{Cox_Forshaw}
B. Cox and J. Forshaw, Double diffraction dissociation at high $t$, 1998,
  hep-ph/9805206.

\bibitem{frankfurt_dvcs3}
L. Frankfurt, A. Freund, and M. Strikman, Diffractive exclusive photon
  production in DIS at HERA, 1997, hep-ph/9710356.

\bibitem{frankfurt_dvcs1}
L. Frankfurt, A. Freund, and M. Strikman, Deeply virtual compton scattering at
  HERA - a probe of asymptotia, 1998, hep-ph/9806535.

\bibitem{frankfurt_dvcs2}
L. Frankfurt, A. Freund, and M. Strikman, DVCS at HERA, 1998, hep-ph/9806406.

\bibitem{Jung}
H. Jung, D. Kr\"ucker, C. Greub, and D. Wyler, Z. Phys. {\bf C} {\bf 60},  721
  (1993).

\end{thebibliography}

% \begin{figure}
% \caption{}
% \label{}
% \end{figure}

% \begin{table}
% \caption{}
% \label{}
% \end{tabular}
% \end{table}

\end{document}